\documentstyle [12pt]  {article}
 \evensidemargin\oddsidemargin

\begin{document}

\title {Nonlinear Quantum  Nonlocality and its Cosmophysical Tests}
\author {{ S.N.Mayburov} \\
 Lebedev Inst. of Physics\\
   Leninsky Prospect 53,  Moscow, Russia, RU-117924\\
  e-mail: mayburov@sci.lebedev.ru \\}
 \date { }
 \maketitle







%
%

%
%


\begin{abstract}

Analysis of  Bell-EPR nonlocal correlations in microscopic measurement theory framework indicates that novel  quantum nonlocality effects can exist. In particular, it can result in  distant correlations between the systems of elementary quantum objects or particles.   Doebner-Goldin  nonlinear quantum formalism applied for  nonlocal correlation  description, comparison with some cosmophysical experiment results discussed.




\end{abstract}



\section {Inroduction}

In 1964 J.S. Bell published the result now known as Bell theorem, which defines the basic properties of quantum-mechanical nonlocal correlations  \cite {Bel}. Nowadays  it was confirmed experimentally and initiated extensive development of quantum information theory and its applications \cite {Ved,Jaeg}. However, despite that tremendous progress, J.S Bell himself and  other researchers argued that quantum nonlocality phenomena are far from complete understanding and can possess some unknown features \cite {Bel,Nor}. In the same vein, many alternative nonlocality theories, prompted by similar thinking, were proposed \cite{Bed,Gil,Cra,Kor,Val,Kup,Sen,Wale}. Additional arguments in favor of these doubts provide the analysis of Bell-EPR correlations from the point of microscopic measurement theories \cite {Bed,Gil}. 
  The problem of quantum measurement or wave function collapse discussed extensively for long and reviewed in many books and papers \cite {Busch, Desp,Bas}.
  Dynamics  of quantum mechanics (QM) described by the Schroedinger equation, which is linear and local \cite {Sud,Sak}. In its framework, the measurement of quantum observable can be treated also as the  interaction of measured state and measuring device (detector) \cite {Desp,Bas}. However, in QM axiomatic its final state postulated to follow the independent projection postulate, which corresponds to nonlinear and stochastic measurement outcome. In early QM days the projection postulate necessity  was advocated by macroscopic nature of measuring apparatus, which supposedly should make its evolution classical (Heisenberg cut)\cite {Desp}. Meanwhile,  multiple experiments in microscopic and mesoscopic domains  contradict directly to this assumption, no deviations from linearity were found even for relatively large objects \cite{Scu,Wal}.  Plainly, detectors consist of elementary objects 
	$-$ particles, atoms, etc., hence it’s seems reasonable to consider the measurement as the sequence of multiple interactions of measured object with these detector elements. Different variants of such microscopic interaction models were proposed, yet it follows  that in standard QM framework,  even for very large number of detector constituents or its environment elements, with which they can interact (decoherence), the resulting system evolution still will be linear, and thus no stochastic measurement outcomes can appear \cite {Busch,Desp}. Therefore, it was proposed that   at fundamental level quantum evolution equation should include also small nonlinear or stochastic terms, which would dispatch in sum the resulting stochastic outcomes \cite {Gil,Bas}.           
  
Such measurement formalism  becomes even more complicated if to consider also the measurement of Bell-EPR correlated state pairs or more complicated correlated systems. Suppose that parameters of two Bell-EPR correlated particles $\nu_{1,2}$ are measured  by corresponding detectors $D_{1,2}$, being divided by space-like interval. Then, as follows from 
Bell theorem,  measurement outcomes of  some $\nu_{1,2}$ observables will be statistically correlated \cite {Gil,Desp}. In this framework, it demands that the elementary interactions of $\nu_{1,2}$ with $D_{1,2}$ elements should be distantly correlated, i.e. entangled.  
Detailed analysis of such NC dynamical mechanism was performed by E.J. Gillis, here we mainly follow its context \cite {Gil}. It supposes that individual pairs of elementary interactions in $D_1, D_2$ are entangled and being accounted  over whole detector volumes constitute complicated and indeterministic event sequence. In this framework, such fundamental interactions of $\nu_{1,2}, D_{1,2}$ element pairs are  nonlinear and nonlocal. Despite that this model still isn' t  finalized due to its formalism complexity and some  open questions  exist, obtained results indicate that consistent microscopic measurement theory can be constructed. Hence it gives hope that collapse problem can be resolved within standard physics realm without addressing to such radical resorts as many world interpretation or subjective collapse by human brain \cite {Desp2,Ev}. Information-theoretical analysis of 
quantum measurements supports its feasibility \cite {May3}.  
  
One of this approach problems concerns with hypothetical correlations between distant individual pairs  of interacting elementary objects. Really, if such correlated evolution  is universal for quantum dynamics, then such nonlocal correlations (NC) between distant elementary object interactions can exist, in principle, not only in the measurement processes, which  incorporate many such elementary object pairs, but also in arbitrary quantum processes like particle scattering or  system decay. Possible existence of such microscopic NC effects was hypothesized earlier,   basing  on  astrophysical arguments \cite{Val,Kor}. In this paper we consider  nonlinear model of such distant microscopic  correlations, which accounts some microscopic measurement model features. Experimental results  which supposedly indicate such NC effect existence reviewed and compared with  model predictions. First results  for simplified variant of this model published in \cite {May4}.

In our approach this microscopic NC mechanism  supposedly can retain some essential features of standard  Bell-EPR correlations. First of all, it's notable that in Bell-EPR set-up the correlations appear between internal states of distant localized systems, in our our example, these are $D_{1,2}$ internal states. Their eigenstates for measured observable correspond to certain measurement outcomes. Such state correlation appear during particle-detector interactions, which change the measurement system state dramatically, assuming that  it's in  pure state it follows that $|<\psi_{in}|\psi_f>|<1$ for its initial and final states \cite {Desp}.
It indicates that in Bell-EPR effect  the evolution of one measurement system supposedly influences nonlocally  evolution parameters  of other distant one and vice versa, as the result,   their measurement outcomes will be correlated even if they are separated by space-like interval \cite {Nor}.
There is no energy, momentum or orbital momentum transfer between distant systems during this correlation formation.   

  Basing on that Bell-EPR correlation features let's discuss  the conditions to which hypothetical  microscopic NC mechanism should obey.  Plainly, beside causality demands, such microscopic NC effects should agree with all standard invariance principles, i.e. time, space shift and rotation symmetries. In accordance with considered Bell-EPR properties,  it assumed that NC by itself $\rm can't$ transfer the energy, momentum or orbital momentum  between distant objects, such transfer can be performed by conventional fields only.
 It will be shown 
that exploit of nonlinear NC Hamiltonians permits to change system states even for such constraints. 
 Let's discuss which quantum systems can be most sensitive to such  NC influence. Consider two distant systems  $S_1, S_2$ evolving during time interval $\{t_0, t_f \}$.  
From  these assumptions it follows that according to QM rules  the initial  $S_1, S_2$ states can' t be stationary and non-degenerate, because such states are ground ones and
				possess  the minimal possible energy, and only some essential energy transfer can make them to evolve to another ones which are excited states.
Hence the only reasonable possibility is that $S_1, S_2$   are degenerate systems, i.e. they have several states with the same energy and  evolve from its initial state to another degenerate one.
  Example of such system  is the particle at energy level $E$ confined in symmetric double well potential divided by potential wall of the maximal height $U_m$ such that $E< U_m$. 
	 Suppose that  system  $S_1$ has two degenerate orthogonal states  $g_1, g_2$ in these wells and at $t_0$  it is in the state $g_1$ confined in one well.
	Thereon,  due to under-barrier tunneling it would spread gradually into other well  \cite{Sak,Per}, so that it will evolve with the time  to some $g_1, g_2$  superposition. In this case, hypothetical $S_2$ NC  influence on $S_1$, in principle, can change the final state parameters, in particular, resulting  $g_1, g_2$  probabilities at $t_f$. Due to reciprocal $S_1$ influence,  analogous $S_2$ evolution  perturbation would occur, if $S_2$ possesses similar structure. If this is the case, $S_1, S_2$  measurements  would indicate deviations from QM predictions.  Such state degeneration is typical for many molecular and nuclear systems, in this paper nonlinear model of analogous NC processes will be considered, underlying mathematical  formalism was formulated in \cite {May}. 

         \section {Experimental Indications}

Here we review experimental results which supposedly evidence for microscopic NC existence and can prompt  possible model features.
		Up to now it was acknowledged that due to strong nuclear forces no environment influence can change  decay parameters of unstable nuclei significantly \cite {Mar}.  However,  recent results indicate that some cosmophysical factors related to Earth motion along its orbit and solar activity can influence them, in particular, their life-time and decay  rate  \cite {Bog,Fis,Al,Al2,Alb}.
 First results, indicating   deviations from standard exponential $\beta$-decay rate
  dependence, were obtained during the precise measurement of
$^{32}$Si isotope decay rate \cite {Alb}. In addition to standard decay exponent, sinusoidal annual
oscillations with the amplitude about $6*10^{-4}$ relative to total
decay rate and maximum in the end of February, were
found. Since then,
  the annual oscillations of
 $\beta$-decay rate for different heavy nuclei from Ba to Ra
 were reported,
for most of them the  oscillation amplitudes are of the order
$5*10^{-4}$ with their maximum on the average at mid-February \cite
{Fis}. 
  Life-time of  $\alpha$-decay isotopes
$^{212}$Po, $^{213}$Po, $^{214}$Po was measured directly, the annual and daily
oscillations with amplitude of the order $7*10^{-4}$, with annual
minimum at March  and daily minimum around 6 p.m. were found
during 6 years of measurements \cite {Al,Al2}.
 It was shown also that decay rates of 
$^{53}$Mn, $^{55}$Fe $e$-capture and $^{60}$Co $\beta$-decay correlate  with solar activity, in particular, with intense solar flare moments, preceding them for several days; in this case, observed decay variations 
are of the order 10$^{-3}\quad$  \cite {Bog,Fis}. 

 Parameters of some
chemical reactions also demonstrate the similar dependence on solar activity and periodic cosmophysical effects \cite {Pic,Shn,Tro}. First results were obtained for bismuth chloride hydrolysis, its reaction rate was shown to correlate  with solar Wolf number and
intense solar flare moments \cite {Pic}.  It was demonstrated that for biochemical unithiol oxidation  reaction its rate correlates with solar activity, in particular, with intense solar flares
and it also grows proportionally to Wolf number. Besides, it was found that its rate correlates  with periodic Moon motion and Earth axis nutation \cite {Tro}. 
Takata biochemical  blood tests  also indicate strong  influence of solar activity and Sun position on its results \cite {Tak}. It performed via  human blood reaction with sodium carbonate Na$_2$CO$_3$ resulting in blood flocculation. Its  efficiency parameter  demonstrates fast gain for the blood samples taken  6-8 minutes before astronomic  sunrise moment, this gain continues during  next hour.  Its value demonstrates approximate invariance during the solar  day and gradual decline  after sunset, such daily dependence conserved even in complete isolation from electromagnetic fields and solar radiation. Such parameter behavior is independent of mountain or cloud presence, which can screen the Sun during solar day. This parameter also rises with solar  Wolfe number growth and for  test location shift in the direction of Earth  equator. Significant  parameter correlations with  Sun eclipse moments also were observed; some of  these results were confirmed  by other researchers \cite {Kaul,Gie,Jez,Kol,Sar}.

Testing blood samples taken  at different altitudes up to several km, authors concluded that influence source isn 't the Sun itself, but Earth atmosphere at  altitudes higher than 6 km \cite {Tak,Kie}. $\rm It's$  established now that during solar day  intense photochemical reactions occur at such altitudes, in particular, O$_2$, SO$_2$, NO$_2$ molecule destruction by ultraviolet solar radiation, which results in ozone and other compound synthesis  \cite {Mc}.  Hence it can be supposed that those photochemical reactions induce NC influence which changes the results of blood reaction with Na$_2$CO$_3$. It can explain, probably, why  reaction rate gain   starts 6-8 min. before
astronomic sunrise and is independent of cloud or mountain presence. Really, at that time solar radiation already reaches Earth atmosphere at such altitudes, and so photochemical reactions occur there,  clouds or mountains $\rm cannot$ absorb this radiation, being located  at lower altitudes.  
Similar daily variations of deuterium diffusion rate into palladium crystal also were reported \cite {Lenr}.
       
 Experiments of other kind also exploit  biochemical and organic-chemical reactions, example is reaction of ascorbic acid with dichlorphenolindophenol \cite {Shn,Shn2}. Authors noticed first that  dispersion of their reaction rates can change dramatically from day to day, sometimes by one order of value, whereas  average reaction rate practically $\rm doesn 't$ change   \cite {Shn,Shn3}. Further studies have shown connection of this effect with some cosmophysical factors, like solar activity, Wolf numbers, solar wind and orbital magnetic field. In particular, average  rate dispersion becomes maximal during solar activity minima of 11 year solar cycle \cite {Shn,Shn4}.  
 Shielding of chemical reactors from external electromagnetic field in iron and permalloy boxes  practically $\rm doesn'$ t change the reaction  dispersions, hence such cosmophysical influence $\rm cannot$ be transferred by  electromagnetic fields.

   Individual nuclear decay or chemical reaction acts normally are independent of each other \cite {Mar}; such stochastic processes are  described by Poisson probability distribution \cite {Korn}.  For this distribution, at any time interval $dT$  the dispersion of decay count number  $\sigma_p=N^\frac{1}{2}$ ,   where  $N$ is average count number per $dT$. 
If  resulting dispersion $\sigma < \sigma_p$, it means that this 
 process  becomes  more regular and  self-ordered and described by
	sub-Poisson statistics with corresponding distribution \cite {Man}.
For $\sigma \to 0$  time intervals between events tend to be constant. If on the opposite  $\sigma_p<\sigma$, it corresponds to super-Poisson statistics, which is typical for collective chaotic processes; in both cases it can be supposed that solar activity acts on reaction volume as the whole, similarly to crystal lattice excitations. For quantum systems such dispersion variations  are typical for  squeezed states \cite {Man}, such approach will be exploited here. 
These results evidence that high solar activity makes  molecular
systems to perform chemical reactions in more self-ordered and regular way. 
In general, similar considerations are applicable to arbitrary statistical distributions
of studied  systems not only Poisson-like ones. 
$\rm It$ is notable that similar temporary Bell-EPR correlations appear in 
 unstable system decays, example is electromagnetic $\pi^0$-meson decay: $\pi^0 \to 2\gamma$ \cite {Mar,Sud}. In that case, if both $\gamma$-quanta detected at equal distances from decay point, then detection moments $t_{1,2}$ in two detectors are correlated according to probability distribution  
$$
       p^d
			(t_1,t_2)=\frac{1}{\sqrt{\pi}\sigma_t}\exp-[\frac{(t_1-t_2)^2}{\sigma_t^2}]
$$
 where $\sigma_t$
 is quantum uncertainty of event time measurement \cite {Sud}. It means that for such decay event ensemble the photon statistics will be sub-Poissonian, corresponding to considered  event distribution for $\sigma < \sigma_p$. Similar effect was observed also for optical photon down-conversion
\cite {Man}.
 Summing up described results, they assume  possible existence of  distant interactions of nonelectromagnetic origin, which supposedly can be attributed to hypothetical NC effects.
Up to now no alternative consistent explanations of such distant influences were proposed.  
 



        \section {Microscopic nonlocality model}

 
        Experimental results considered here evidence that these  distant  correlations occur for evolving quantum systems when such system by itself suffers significant transformation even without hypothetical NC influence, in particular, it occurs for chemical  reactions and decays of unstable nuclei.
In standard QM framework, system evolution operator defined mainly by quantum-to-classical correspondence, for assumed NC effects such guidelines are absent, here it will be constructed  basing only on general QM principles and cited experimental results. 
 We shall consider  NC effects for nucleus $\alpha$-decay, which supposedly  were
observed in experiments described in \cite {Al,Al2}.  
   Gamow theory of  $\alpha$-decay
assumes that in initial nucleus state, free  $\alpha$-particle
already exists inside the nucleus, however, its  energy $E$ is
smaller than  maximal height of potential barrier $U_m$  constituted by
nuclear forces and Coulomb  potential \cite {Gam}. Hence
$\alpha$-particle can leave nucleus volume only via quantum
tunneling through this barrier.
Therefore, alike for considered double well example,   $\alpha$-particle energy is the same inside and outside nucleus, and  corresponding inside-outside states   are degenerate.
  Hence such  degeneration permits, in principle, for  hypothetical NC mechanism to change nucleus decay rate without any energy transfer to $\alpha$-particle, but just changing the barrier transmission rate. 
 Below  NC effects will be considered for system $S$ of $N$ independent, identical nucleus $\{A_i\}$, its initial state is $\{A_i\}$ product state.
In Gamow model $\alpha$-particle Hamiltonian  for metastable nucleus
\begin {equation} 
                    H=\frac{\vec{P}^2}{2m}+U_N(\vec{r})     \label {X0}
\end {equation} 
 where $m$ is $\alpha$-particle mass, $\vec P$ is its momentum operator, $U_N$ is nucleus barrier potential, $\vec r$ is $\alpha$-particle coordinate relative to  nucleus center \cite {New}. If at $t_0$  for arbitrary  $A_i$ $\alpha$-particle was in initial state  $\psi^i(t_0)=\psi^i_0$, then  Shroedinger equation solution for  $\psi^i(t)$ in WKB approximation \cite {Sud,Gam} gives the decay probability rate at given  $t$
 \begin {equation} 
                   p_i (t) = \lambda \exp[-\lambda (t-t_0)] 
											                                        \label {X1}           
 \end {equation}
here $p_i$ is the  time derivative  of $A_i$ total decay probability from  $t_0$  to $t$;  resulting  nucleus life-time is proportional to  $\lambda^{-1}$.
 Hence at $t\to \infty$ $A_i$  state evolves to final state $\psi^i_1$, such that  $<\psi^i_0|\psi^i_1>=0$.
 In fact,  similar considerations
are applicable to the evolution of arbitrary metastable system, like atoms or molecules,
 yet for  $\alpha$-decay its description is most simple \cite {Gam,New}.
In particular, analogous metastabilty properties  are characteristic also for considered chemical reactions \cite {Pic,Tro,Shn,Shn2}. 

   It was argued above that for Bell-EPR correlations  the  evolution of one measurement system influences the evolution parameters of other distant one and vice versa \cite {Nor,Gil}, illustrative example is Bell-EPR temporary correlations considered in previous section. One can suppose that similar mechanism defines the properties of discussed microscopic NC effects. In this framework,
     our   assumption is that  intensity of  NC effect induced by some system $S_1$ will be proportional to some function of $S_1$  transition rate from its initial internal state $\psi_{in}$ to final one $\psi_f$, such that $<\psi_{in}|\psi_f>=0$. Analogously to Bell-EPR correlations, since NC influence should be reciprocal, it reasonable to suppose that such $S_1$ influence on some $S_2$ system
would change  $S_2$  transition rate, and vice versa for $S_2$ NC influence on $S_1$.
 In particular,  it can be supposed arbitrarily  that NC influence intensity of  nucleus $A_i$ decay on the evolution of another nucleus is proportional to $p_i(t)$ of eq. (\ref{X1}). 
		This assumption will be reconsidered below in QM formalism framework, it will be shown that it’s applicable only for some approximation and in general NC influence  described  by corresponding QM operator. 
			
 Experimental results discussed in previous section indicate that NC  influence can make collective system evolution less chaotic and more regular, in particular, can result in squeezed states with sub-Poisson statistics.  It’s notable that  self-ordering  is quite general feature of quantum dynamics, examples are crystal lattices or atomic spins in ferromagnetic.
	 Besides the collective system self-ordering,  other forms of   symmetrization,  induced by  standard QM dynamics, concerned with individual object state. Example is elastic particle scattering, in that case,
the final state possesses larger	angular symmetry than incoming plane wave state \cite {Sak}.
These analogies together with cited experimental data permit to suppose that  such NC influence  transforms system evolution so that it results in more time-symmetric and self-ordered states, in comparison with its nonperturbed evolution, our choice of NC Hamiltonians will be prompted by this assumption.
	For example,  enlargement of metastable system life-time  can be  treated as  the growth of evolution symmetry, because 
the decay probability rate  $p_c(t)$ becomes more homogeneous in time, its asymptotic  limit is $p_c(t)\to const$ for $t>t_0$.
 $\rm It$ is notable that experimental results reviewed above demonstrate  enlargement  of nucleus life-time induced by enhanced solar activity \cite {Fis,Bog}, the same is true for its influence on some chemical reaction rates \cite {Pic,Tro}. 

Typical experimental accuracy  of nuclear decay time moment measurement $\Delta t$ is several nanoseconds \cite {Mar}. Formally, such measurement  described as the sequence of multiple consequent state measurements divided at least by $\Delta t$ interval. If first one shows that nucleus  is in $\psi_0$ state, and next one that it’s in the state $\psi_1$, it means that nucleus decay occurred during this time interval \cite{Sud,Sak}.         
       In QM formalism, a general state of quantum system $S$ described by density matrix $\rho$, if  $A_1, A_2$ nuclei are $S$ components, the partial $A_{1,2}$ density matrices  $\rho_{1,2}$ can be defined.  For each $A_i$ it turns out that if some other $S$ components are also measured, then its decay probability rate
			would differ
			from eq. (\ref{X1}) and becomes
\begin {equation}			
                  \gamma_i (t)=\frac{\partial}{\partial t} Tr \rho_i (t)P_1^i    \label {X3}
\end {equation}                                                              
where $P_1 ^i$ is projector on $A_i$ final state \cite {Sud}.




     \section {Nonlinear decay formalism}

It was supposed that  NC effects should not change the system
average energy, however, if corresponding NC Hamiltonian is  linear
operator then for $\alpha$-decay in Gamow model  this condition is violated \cite {May}. It will be shown here that nonlinear Hamiltonians can satisfy much better to this condition.
It's acknowledged now  that   nonlinear corrections to standard
QM  can exist  at fundamental level \cite
{Wn,DG,DG2}. 
 In nonlinear QM formalism, particle evolution described by nonlinear
Schroedinger equation of the form 
\begin {equation}
   i\hbar \partial_t \psi=   -\frac{\hbar^2}{2m}\bigtriangledown^2\psi +U(\vec{r},t)\psi
    + F(\psi,\bar\psi)\psi  \label {AXA}
\end {equation}
where $m$ is particle mass, $U$ is system potential, $F$ is
arbitrary functional of system state.
Currently, the most popular and elaborated  nonlinear QM model is by Doebner and Goldin (DG)  \cite {DG,DG2}. In
its formalism,  simple variant of nonlinear term is $F= \frac{\hbar^2 \Gamma}{m}\Phi$
where \begin {equation}
    \Phi= \bigtriangledown^2 + \frac{|\bigtriangledown\psi|^2 }{|\psi|^2}  \label {BXB}
\end {equation}
is nonlinear operator, $\Gamma$ is real or imaginary  parameter which, in principle, can depend on time or other external factors, here only real $\Gamma$ will be exploited. 
 With the notation
 \begin {equation}
    H^L=-\frac{\hbar^2}{2m}\bigtriangledown^2 +U(\vec{r},t) \label {BB}
\end {equation}
 we abbreviate eq. (\ref{AXA}) to
 $i\hbar{\partial_t\psi}=H^L\psi+F\psi$ where in our case, $H^L$ is Gamow Hamiltonian.


Main properties of eq. (\ref{AXA})  were
studied in \cite {DG,May}, for constant $\Gamma$ they can be  summarized  as follows:
(a) The probability is conserved. (b) The equation is homogeneous.
(c) The equation is Euclidean - and time-translation invariant 
for $U=0$.
 (d) Noninteracting particle subsystem remain
uncorrelated (separation property).
 Distinct values of $\Gamma$
 can occur for different particle species.
 (e) For $U=0$, plane waves $\psi=\exp[i(\vec{k_0}\vec{r}-\omega t)]$
with $\omega=E/\hbar$, $|\vec{k}_0|^2=2mE/\hbar^2$ are solutions
both for real and imaginary $\Gamma$.
(f) Writing $<Q>=\int
\bar{\psi}\hat{Q}\psi d^3x$ for operator expectation value,  since
 $ \int \bar{ \psi}F\psi d^3x=0 $ for arbitrary $\psi$, the energy functional for 
 solution of eq. (\ref{AXA}) is $<i\hbar \partial_t>=<H^L>$. Hence the average system energy   would change insignificantly if not at all if $F$ added to initial Hamiltonian, therefore, it
advocates DG ansatz application in NC models.  In particular, it will be shown that in WKB approximation, which is the main ansatz for decay calculus, the   energy expectation value doesn't change in the presence of such nonlinear term.

As was noticed, Bell-EPR correlations appear between internal states of distant systems,
by the analogy, we suppose that in our model nonlinear term $F$ acts on $\{A_i\}$ nuclei internal states $\psi^i(t)$.  
It's notable that nonlinear term $F$ in $\alpha$-particle Hamiltonian  can modify 
the particle tunneling rate through the
potential barrier. In particular, analytic solution of this problem
was  obtained for rectangular potential barrier, in that case, the barrier transmission rate depends exponentially on $\Gamma$ \cite {May}.  
 To calculate corrections 
 to Gamow model for arbitrary potential $U$,
 WKB approximation for nonlinear Hamiltonian of (\ref{AXA})
 can be used \cite {Sak}. 
 In this ansatz, for rotation-invariant $U$  $\alpha$-particle wave function  reduced
to $\psi=\frac{1}{r}\exp(i\sigma(r)/\hbar)$; function $\sigma(r)$
can be decomposed in $\hbar$ order $\sigma=\sigma_0+\sigma_1+...$, here $r=|\vec{r}|$ is the distance from nucleus center
\cite {Sud,Sak}.
 Given
$\alpha$-particle with energy $E$, one can find the distances $R_0,
R_1$ from nucleus center at which $U(R_{0,1})=E$. Then, for our
nonlinear Hamiltonian the resulting equation for $\sigma_0$ 
\begin {equation}
       (\frac{1}{2m}-\Lambda)   (\frac {\partial \sigma_0}{\partial {r}})^2=E-U(r)  \label {AZZ}
\end {equation}
where $\Lambda=\frac{2\Gamma}{m}$ for $R_0\leq r\leq R_1$, $\Lambda=0$ for
$r<R_0$, $r>R_1$ \cite {May}. Its solution for $R_0\leq r\leq R_1$ can be
written as
\begin {equation}
   \psi(r)=\frac{1}{r}\exp(i\sigma_0/\hbar)=\frac{C_r}{r}\exp[-\frac{1}{\hbar}\int\limits_{R_0}^{r} q(\epsilon)  d\epsilon]        \label {ZX9}
\end {equation}
where $C_r$ is normalization constant,
 \begin {equation}
   q(\epsilon)=\{\frac{2m[U(\epsilon)-E]}{1-4\Gamma}\}^{\frac{1}{2}}
      \label {ZX10} 
\end {equation}
 Account of higher order $\sigma$ terms practically
doesn't change
 transmission coefficient which is equal to
\begin {equation}
  D=\exp[-\frac{2}{\hbar}\int\limits_{R_0}^{R_1}q(\epsilon)
   d\epsilon]=\exp[-\frac{\phi}{(1-4\Gamma)^\frac{1}{2}}] \simeq
   \exp[-\phi(1+2\Gamma)]  
    \label {ZXX}
\end {equation}
here $\phi$ is constant for given nucleus, whereas $\Gamma$, in principle, can change in time, assuming that its time change scale is much larger than the barrier transition time. Note that $\Lambda$ term induced by nonlinearity  doesn't change the average particle energy in comparison with corresponding  linear ansatz $H^L$. 
To calculate nucleus life-time, $D$  multiplied by the
number of $\alpha$-particle kicks  into nucleus potential wall per
second $n_d$, so it gives $\lambda=n_d D$ \cite {Gam},   for DG model $n_d$ doesn't depend
on  $F$ term  \cite {May}.  



It is natural to assume that NC effect  for any system  of restricted size grows with the number of system constituents $N$ involved into reactions. 
For the case of two  systems of which one of them $S_1$ is large and other one $S_2$ is small,  for them NC effects supposedly realized in master regime, i.e. $S_1$ can significantly influence $S_2$  state and make it evolution more ordered and symmetric  as well as its own evolution, whereas $S_2$ practically $\rm doesn’t$ influence  $S_1$ state evolution. 
  It can be assumed also that in this case, $S_2$  self-influence  NC effect is insignificant in comparison with $S_1$ NC influence.
 Then, resulting NC effect in $S_2$  supposedly depends on  $S_1$ evolution properties and $S_1, S_2$ distance $R_{12}$.
It was argued that enlargement of nucleus life-time, i.e. $p_i(t)\to const$, can be interpreted as system evolution symmetrization. 
Let’s study  how such NC influence can be described in master regime approximation.
Consider  two nucleus systems $S_1 ,S_2$   with the average distance $R_{12}$ between $S_1, S_2$  elements,  supposedly it's much larger than $S_{1,2}$ size; for the simplicity we'll consider only static situation when object positions are fixed. $S_1$  is  the set  of  $N_1$  unstable nuclei $\{A_l\}$ prepared at  $t_0$  with decay probability rate described by eq. (\ref{X1}).  $S_2$   includes just one unstable nucleus $B$ prepared also at  $t_0$, its evolution normally described by Gamow Hamiltonian $H^L$  ansatz of (\ref{BB}). Its decay constant $\lambda_b$, in principle,  can differ from $\lambda$  of eq. (\ref{X1}) if for $B$ 	$U\ne U_N$ of eq. (\ref{X0}).  In such set-up,  presumably  NC effects induced by $S_1$ would influence  $B$ evolution and perturb also  its own evolution as well. Hence for $S_1$ nuclei their initial decay probability rate can change to some  $p^v_i(t)$.  Suppose that all geometric factors of such NC influence on $B$ for given $S_1$ described by  real function  $\chi(N_1,R_{12})$  which absolute value grows with $N_1$ and diminishes with $R_{12}$, i.e. $\chi$ is phenomenological NC propagation function.  Resulting   corrections to $H^L$ are supposed to be small and so can be accounted only  to the first order of  $\Gamma$.  Basing on assumptions discussed above, in particular, that resulting NC effect proportional to $S_1$ total transition rate, it follows that parameter $\Gamma$ in $F$ nonlinear term becomes the function  $\Gamma(R_{12},t)=\chi(N_1,R_{12}) p^v_i(t)$, so that phenomenological  $B$ Hamiltonian 
		\begin {equation}
                H^d (t)=H^L+\frac{\hbar^2}{m} \chi(R_{12}) p^v_i (t)\Phi	\label {BZ}
\end {equation}                                         
where $\Phi$ is nonlinear operator of eq. (\ref {BXB}) for $B$
 $\alpha$-particle. 
 Solving evolution equation in WKB approximation it follows from eq. (\ref{ZXX}) that if no measurements of $S_1$  states were performed, then 
 $B$ decay probability rate
\begin {equation}
                          p_b'(t)=C_b \exp\{-[(t-t_0)\xi(t)]\}    \label {BZ2}
\end {equation}																						
here $C_b$  is normalization constant
\begin {equation}
   \xi(t)=\lambda_b(\frac{\lambda_b}{n_d})^{2\Gamma(R_{12},t)}  \label {BZ3}
\end {equation}
Hence for such ansatz, $B$ decay probability rate $p_b'(t)$  depends on total $S_1$ nuclei decay probability rate
and would differ from $B$ probability rate $p_b(t)$ in Gamow model; for $\chi>0$ $B$ life-time would enlarge.
 It can be  assumed  that $p^v_i(t) \approx p_i(t)$  of eq. (\ref{X1}), because in
our model typical $S_1$ NC self-influence expected to be small.  
Note that $B$ kinematic parameter $n_d$ practically doesn't change in this case \cite {May}. 

     Now  NC effects between  elementary systems will be considered beyond master regime approximation. Suppose that for system $S_1$  $N_1 =1$ and  nuclei $A_1, B$  states described by wave functions $\psi^1(t), \psi^b(t)$ correspondingly.	Then for the same initial conditions as above, the system initial wave function $\psi_s=\psi_0^1 \psi_0^b$. 
In QM framework, probability of given state $\varphi$ presence at particular time defined by its projector $P_{\varphi}$, in accordance with it, $A_1$ transition rate described by the operator $Q^1_1=\frac{dP_1^1}{dt}$,  where  $P_1^1$ is projector on $A_1$ final state of eq. (\ref{X3}),  $Q_1^b$ is corresponding derivative for projector on $B$ final state; they can be calculated from Erenfest theorem \cite {Sak}.   Therefore $\Gamma$ for $A_1, B$ terms replaced by the operators $\Gamma_{A}=\chi(N_1,R_{12})Q_1^b$ and $\Gamma_{B}=\chi(N_1,R_{12})Q_1^1$,  corresponding $A_1, B$ Hamiltonian	
\begin {equation}	
   H_s (t)=H^L+ \frac{\hbar^2}{m}\chi(N_1,R_{12} )Q^1_1\Phi +H_1+ \frac {\hbar^2}{m} \chi(N_1,R_{12} ) Q_1^b \Phi_1       \label {Z2Z}
\end {equation}											where $\Phi_1$  is nonlinear operator of eq. (\ref {BXB}) for $A_1$, $R_{12}$ is $A_1, B$ distance.
 It follows that in first perturbation order  $<Q_1^{1,b}>=p_{1,b} (t)$       
with $p_1(t)=p_i(t)$ of eq. (\ref{X1}). Since $A_1, B$ operators commute, transition rate operators can be replaced by their expectation values  
\begin {equation}
   H_s (t)=H^L+\frac{\hbar^2}{m} \chi(N_1,R_{12}) p_1 (t)\Phi+H_1+\frac{\hbar^2}{m} \chi(N_1,R_{12}) p_b (t)\Phi_1         \label {ZZZ}
\end {equation}                                         
 Solution of evolution equation for  Hamiltonian of eq. (\ref{Z2Z}) gives transition rate for $B$ decay  described by eq. (\ref {BZ2}) with $p^v_i(t)=p_1(t)$, for $A_1$ resulting rate $p’_1(t)$ can be  calculated  analogously. Hence obtained ansatz supports the use of eq. (\ref{BZ}) for  NC influence calculations  in master regime. 
Any alternative NC effect ansatz would demand  the use of more complicated operators, hence the hypothesis of its proportionality to  transition rate operator seems reasonable.  Obtained $A_1, B$ states are correlated but not entangled,  so that the system state $\psi_s=\psi^1(t) \psi^b(t)$ at arbitrary time, however, in the next order their entanglement can appear.

Under NC  influence for $\chi> 0$    resulting $A_1,B$ nucleus life-times   becomes larger than initial one. 
Such $S_1, S_2$ evolution modification can be interpreted as the growth of $S_1,S_2$ evolution symmetries such that resulting decay probability rates $p’_1(t), p’_b(t)$  becomes more homogeneous in time in comparison with
 initial $A_1, B$  probability rates.
 It can be supposed also that inverse process, i.e. $A_1$ nucleus synthesis via reaction of $\alpha$-particle with remnant nucleus  would induce the opposite NC effect on $B$,  reducing $B
$ nucleus life-time and so reducing its evolution symmetry.
Hence proposed NC mechanism can change, in principle, the evolution symmetry in both directions enlarging or reducing it.
 Thermonuclear reactions in the Sun
result in production of unstable isotopes \cite {Mar}, hence according to that model, variations of such reaction intensity 
 can result in  variation of solar NC influence rate on nuclear decay parameters on the Earth \cite {Fis,Bog}. Such reaction rate variations  supposedly  can occur during solar flare formation, because it results in intense  ejection of charged particles and $\gamma$-quanta from solar surface \cite {Han}. It supposedly can be the reason for observed correlations between solar flare moments and isotope decay rate decline on Earth \cite {Fis,Bog}.
 It was assumed above that NC should not change any subsystem  average energy,  in our  model this condition fulfilled in WKB approximation. System momentum and orbital momentum conserved due to Hamiltonian rotational symmetry.

		



\section {Squeezed State Production}

		Now let's consider NC model, which describes decay self-ordering symmetry
	gain, resulting in sub-Poisson event statistics. For that purpose multiple time formalism for evolution operator calculation will be used, which is standard approach for time-dependent Hamiltonians \cite {Sak}.
		  Consider the system $S$  of  $N$  nuclei,
 as was supposed, due to conjugal  NC influence between $S$ elements, its evolution would become more regular and self-ordered without significant life-time change. Hence $S$ evolution can differ from the case of independent nuclei and would result in the temporary correlation between   decays of $S$ nuclei.
Consider the simplest case $N=2$  with $A_1, A_2$ nuclei prepared at $t_0$ at the distance $R_{12}$.
   $S$ evolution operator can be chosen  analogously to the one for squeezed photon production  in atomic resonance fluorescence \cite {Man}.  In its simple variant, photon production  is suppressed if the time interval between two consequently produced photons is less than some fixed  $\Delta T$. Due to it,  resulting photon production  becomes more regular, and their statistics would become sub-Poissonian.
Suppose that $A_1$ NC influence rate  on $A_2$  characterized by real function  $k(R_{12})$, its absolute value supposedly diminishes as  $R_{12}$ grows, the same function describes $A_2$ NC influence on $A_1$; $k$ can be regarded as NC coordinate Green function. Analogously to  our previous considerations, suppose that $A_1$ NC influence intensity on $A_2$ evolution is proportional to $A_1$ transition rate operator and vice versa, but integrated over some time interval. In this case,  analogously to eq. (\ref {Z2Z})  $\Gamma$ of eq. (\ref{BXB}) becomes the operator. For the simplicity, we assume that $A_1, A_2$ decay  evolution ansatz  can be factorized into $A_1, A_2$ terms. For example,  phenomenological  $A_2$ Hamiltonian 
\begin {equation}	
							H_2^c (T)=H_2+\frac{\hbar^2}{m} \int\limits_{t_0}^
									{T} k(R_{12})\varphi(T-t)Q^1_1\Phi_2dt 	\label {Z1}
\end {equation} 
							 $Q^1_1$ operator defined above, $\Phi_{1,2}$ are $A_{1,2}$ nonlinear operators
	of eq. (\ref{BXB}) with corresponding notations;
		$\varphi$ is causal Green function	 
		$$
		                            \varphi(\tau)=\eta(\tau-\nu)-\eta(\tau)
		$$
 Thus, corresponding NC time dependence described as the difference of two step functions
 $\eta(\tau) = \{0, \tau<0; 1, \tau \geq 0\}$ which is simple variant of such ansatz
 \cite {Korn}. Here $\nu$  is the time range in which $A_1, A_2$ decay acts are correlated.
	Hence $A_2$ Hamiltonian $H^c_2$ is time-dependent, at given time moment $T$ it depends on $A_1$ decay rate  during time interval $\nu$ previous to $T$.  Analogous modification occurs for $A_1$ Hamiltonian with corresponding index change.  As the result, such NC influence for $A_{12}$ described by evolution operator with multiple time ansatz 	
\begin {equation}
								W(T)=C_t exp\{-\frac{i}{\hbar(T-t_0)} \int\limits_{t_0}^{T} 
								\int\limits_{t_0}^{T} [H_1 (t_1)+H_2 (t_2)
						+(T-t_0)G(t_1,t_2)]dt_1 dt_2\}              \label {Z2}
\end {equation}	
where $C_t$ is time-ordering (chronological) operator \cite {Sak}.
												Third term in this equality is NC  dynamics term, it  suppresses nucleus decays at small time intervals between them, so that
 \begin {equation}
         G(t_1,t_2 )= \frac{\hbar^2}{m}k(R_{12}) [\varphi(t_1-t_2) Q_1^2\Phi_1+
				 \varphi(t_2-t_1)  Q^1_1\Phi_2]  \label {Z23}       
\end {equation}	
where $\Phi_{1,2}$ are of eq. (\ref{BXB}), $Q_1^2$ is $Q_1^1$ equivalent for $A_2$ . Note that the second right-side term corresponds to  $H_2^c$  Hamiltonian of eq. (\ref{Z1}).
Due to  $A_1, A_2$ operator commutativity, the operators $Q^{1,2}_1$ for $A_{1,2}$ in first perturbation order  can be replaced by their expectation values $\gamma_{1,2}(t)$ of eq. (\ref{X3}).
  If no measurement of $A_{1,2}$  state was performed for $t_f <T$, then $\gamma_{1,2}(t)=p_i(t)$ of eq. (\ref{X1}). 
 Otherwise, if such measurement  was done at some  $t_f$  and $A_i$ was found to be in the  final state, then for $t_a >t_f$  it follows   that $\gamma_i( t_a)=0$. 
		If no $A_{1,2}$ measurement was done, then in WKB approximation    the 
 joint $A_{1,2}$ decay probability rate $p_s$ for $k>0$ will differ  from independent decay case when $p_s(t_1,t_2) =p_1(t_1) p_2(t_2)$ and is equal to
   \begin {equation}
         p_s (t_1,t_2 )=C \lambda^{2+4\theta} \exp [-g(t_1,t_2 )(t_1+t_2-2t_0 )] \label {Z99}
    \end {equation}
where $C$ is normalization constant, analogously to eq. (\ref{BZ3})
\begin {equation}
                g(t_1,t_2) =\exp[ (1+2\theta) \ln \lambda]   \label {Y1}
\end {equation}
where $\lambda$ is from eq. (\ref{X1})										
\begin {equation}
 \theta=\frac{\hbar^2}{m} k(R_{12})[\eta(t_1-t_2) \varphi(t_1-t_2) \gamma_2 (t_2 )
+\eta(t_2-t_1) \varphi(t_2-t_1) \gamma_1 (t_1 )]    \label {Z98}
\end {equation}
 Due to it, if the time interval between two decay moments is less than $\nu$,  the  nucleus decay rate will be suppressed, and resulting decay event distribution will become more regular,
i.e. sub-Poissonian.
 Note that in the considered approximation $A_1, A_2$ states are correlated, but not entangled.
 For  $N > 2$ the considered NC dynamics term in $W(T)$ would change to
                                                $G (t_1, . . . ,t_N) dt_1 . . . dt_N$  
with corresponding integration over $N$   independent  time parameters. As the result, for analogous $G$ ansatz  the joint decay probability of two arbitrary consequent decays will be suppressed for small time intervals between them, and $N$ decay event distribution  will be sub-Poissonian. 
 Two considered symmetrization  mechanisms, i.e. life-time enlargement and event sub-Poisson symmetrization, in principle, can coexist and act simultaneously in some systems. Here the system self-ordering NC effect was considered, however, some  distant system $S_m$  also can induce, in principle, analogous NC evolution symmetrization in system $S$, as experimental results evidence \cite {Shn,Shn2}. It can be supposed  that analogous NC effect description is applicable also to  chemical reactions and other atomic and molecular systems. 
 Nonlinear Hamiltonians were used here for NC effect description, however, it’s possible also that in collective systems NC effects at fundamental level can be described by linear Hamiltonians, so that nonlinear QM appears as the corresponding effective theory \cite {DG2}.



\section {Conclusion} \label{sec-5}


         Considered experimental results and theoretical analysis evidence that novel communication mechanism between distant quantum systems can exist. $\rm It’s$
				based on specific form of QM nonlocality,  different from well-known Bell-EPR  mechanism, in particular, its effects supposedly  can be observed even between microscopic quantum systems. 				
				         In this paper, microscopic NC effects  studied for  metastable  quantum systems, namely,  $\alpha$-decay  with  nonlinear NC Hamiltonian.
Such microscopic  NC influence supposedly has universal character, in particular, such nonlocal influences  can exist between the systems of scattering particles.
However, for metastable systems NC effects are expected to be more easily accessible for experimental observation due to their relatively long duration. 
Beside
nuclear decays and chemical reactions, such  microscopic NC effects
 can be observed, in principle, for other systems in which
metastability and tunneling plays important role. This is true, in fact, for  biological systems,
 it was proposed earlier that  long-distance correlations observed inside  living organisms and plants can be induced by QM nonlocality   \cite {Pri,Bis}.
However,  standard Bell-EPR mechanism can not induce such NC effects in dense and warm media, which is characteristic for biological systems. Possible description of such biological effects via microscopic NC influence mechanism
discussed in \cite {May2}.
It's worth to stress that in this nonrelativistic model 
nonlinear Hamiltonian terms describe nonlocal effects only, whereas local interactions are linear. 
 EPR-Bohm paradox and 
Bell inequalities demonstrate that quantum measurement dynamics is essentially nonlocal \cite {Bel,Nor}. However, 
as was argued above, it seems doubtful that dynamics of quantum measurements differs principally
from the rest of QM dynamics, more reasonable is to expect 
 that both of them can be described by some universal  formalism, hence the presence  of nonlocal terms in it would be plausible \cite {Gil, Bed}. 
 It $\rm isn’\, t$ clear whether NC Hamiltonians, exploited here, are suitable for the description of measurement processes also. However, $\rm it' s$  notable that particle detectors are usually the metastable systems, so it is reasonable to  consider such models as possible candidates \cite {Mar}.

Concerning with causality  for NC communications, at the moment it is still possible to assume that such  NC can spread  between systems with velocity of light. But even if this spread is instant,  it is notable that usually superluminal signalling in QM discussed for one bit yes/no communications \cite {Bel,Nor}. In our case, to define the resulting  change  of some  parameter expectation value, one should collect significant event statistics which can demand significant time, so it makes causality violation  quite doubtful possibility.  In addition, NC dependence on the distance between two systems expressed by $\chi$ propagation function can be so steep that it also would suppress effective superluminal signalling. Situation can be similar to QFT formalism where some particle propagators spread beyond light cone, but due to analogous factors, it doesn't lead to superluminal signalling \cite {Blo}. Note that some nonlinear QM models by themselves permit superluminal signalling, but it doesn't  contradict to our nonlocal formalism \cite {May,DG2}.






 












\end {document}